\documentclass[12pt,preprint]{aastex}
\usepackage{url}
\usepackage{multirow}
\usepackage[symbol*]{footmisc}

\newcommand{\sat}[1]{\it\uppercase{#1}\rm}
\newcommand{\fig}[1]{Figure~\ref{#1}}

\newcommand{\speed}[1]{#1 km s${}^{-1}$}

\newcommand{\rsun}[1]{${#1}\,R_\sun$}

\shorttitle{Twin CMEs Associated With One Coronal Blowout Jet} %

\shortauthors{Shen et al.}

\begin{document}

\title{ON A CORONAL BLOWOUT JET: THE FIRST OBSERVATION OF A SIMULTANEOUSLY PRODUCED BUBBLE-LIKE CME AND A JET-LIKE CME IN A SOLAR EVENT}

\author{Yuandeng Shen\altaffilmark{1,2,4}, Yu Liu\altaffilmark{1,4}, Jiangtao Su\altaffilmark{3,4}, and Yuanyong Deng\altaffilmark{3,4}}
\altaffiltext{1}{National Astronomical Observatories/Yunnan Astronomical Observatory, Chinese Academy of Sciences, P.O. Box 110, Kunming 650011, China; ydshen@ynao.ac.cn}
\altaffiltext{2}{Graduate University of Chinese Academy of Sciences, Beijing 100049, China}
\altaffiltext{3}{National Astronomical Observatories, Chinese Academy of Sciences, Beijing 100012, China}
\altaffiltext{4}{Key Laboratory of Solar Activity, National Astronomical Observatories, Chinese Academy of Science, Beijing 100012, China}

\begin{abstract}
The coronal blowout jet is a peculiar category among various jet phenomena, of which the sheared base arch, often carrying a small filament, experiences a miniature version of blowout eruption that produces large-scale coronal mass ejection (CME). In this paper, we report such a coronal blowout jet with high-resolution multi-wavelength and multi-angle observations taken from {\sl Solar Dynamics Observatory}, {\sl Solar Terrestrial Relations Observatory}, and Big Bear Solar Observatory. For the first time, we find that a simultaneous bubble-like and a jet-like CMEs were dynamically related to the blowout jet that showed cool and hot components next to each other. Our observational results indicate that: (1) the cool component was resulted from the eruption of the filament contained within the jet's base arch, and it further caused the bubble-like CME; (2) the jet-like CME was associated with the hot component, which was the outward moving heated plasmas generated by the reconnection of the base arch and its ambient open field lines. On the other hand, bifurcation of the jet's cool component was also observed, which was resulted from the uncoupling of the erupting filament's two legs that were highly twisted at the very beginning. Based on these results, we propose a model to interpret the coronal blowout jet, of which the external reconnection not only produces the jet-like CME but also leads to the rising of the filament. Subsequently, internal reconnection starts underneath the rising filament and thereby causes the bubble-like CME.
\end{abstract}


\section{INTRODUCTION}
There are many kinds of jet phenomena in the solar atmosphere, such as H$\alpha$ surges, EUV jets, X-ray jets, EUV macrospicules, and H$\alpha$ macrospicules. They often have a linear or slightly bent structure, and usually associate with microflares at the base. A standard interpretation for these phenomena of different scales is that a small bipolar emerging flux interacts with its surrounding large-scale opposite-polarity, unipolar open fields, and the subsequent magnetic reconnection between them generates the straight outward heated plasma flow guided by the open field lines \citep{shib92}. In this scenario, the reconnected closed fields are observed as a small bright point (microflare) close to the jet base. In despite the difference among various jet-like eruptions such as the temperature, velocity, and scale, we refer to all of them as coronal jets for convenience in this paper, due to their similar physical mechanism in nature.

Coronal jets are usually associated with flux emergence and cancellation \citep[e.g.,][]{chae99,liu04}, and magnetic reconnection is necessary for the production of a coronal jet \citep{rach10}. In a few cases, coronal jets exhibit obvious unwinding motions around the main axis \citep[e.g.,][]{alex99,pats08,liuw09,shen11b}, as well as in small-scale explosive events \citep[e.g.,][]{curd11}. Such unwinding motions suggest that magnetic helicity stored in the closed bipolar emerging flux is transferred to the outer corona. In addition, many coronal jets show simultaneous cool and hot components next to each other, which are thought to be different plasma flows along different field lines but dynamically connected to each other \citep{canf96,chae99}. Other studies have shown that coronal jets could result in coronal mass ejections (CMEs; \cite{liu05a,liu08,wang98,wang02}), supply mass to form or lead to the eruption of a large-scale filament\citep{liu05b,guo10}, and interact with other coronal magnetic structures and thus lead to a sympathetic CME pair \citep{jian08}. In addition, by using observations from {\sl Solar Dynamics Observatory} (\sat{sdo}), \cite{tian11} suggested that the ubiquitous jet-like outflow in the solar corona is an important possible mass source for the solar wind. In terms of theory, numerical simulations in two-dimensional and three-dimensional have been performed and successfully reproduced many important observable characteristics of the coronal jets \citep{shib86,yoko95,yoko96,miya03,miya04,more08,arch10,pari10}.

On the other hand, statistical researches on coronal jets have been documented in a lots of studies \citep[e.g.,][]{shim96,wang98,yama04,liu08,nist09,moor10}. \cite{liu08} classified the coronal jets in the chromosphere into three types: jet-like type, diffuse type, and closed loop type. The former two types were found to associate with large-scale CMEs, while the latter type was not. Therefore, he suggested that the difference in magnetic field configurations in the low corona determined the later evolution process of the coronal jets. \cite{yama04} found about half of their jet samples had the form of a small erupting loop that appears similar to the eruption of a mini-filament. Furthermore, \cite{nist09} found a few coronal jets characterized by a small loop that elongates from the solar surface, and resembled the large-scale CMEs but on much smaller scales. Very recently, by examining many coronal jets in polar coronal holes using soft X-ray observations from {\sl Hinode} satellite \citep{kosu07}, \cite{moor10} found that there is a dichotomy of the coronal jets, namely the standard coronal jets and the blowout coronal jets. According to their classification, about two-thirds of coronal jets belong to the standard jets, and about one-third are the blowout jets. The most distinctive feature of the blowout jets is that the base arch, which is assumed to be sheared and twisted and often carried a small filament within it, can experiences a miniature version of blowout eruption that is usually evidenced in the large-scale CME eruptions \citep[e.g.,][]{ster04,shen11c}. Furthermore, \cite{ster10} identified that the chromospheric spicules are the counterparts of coronal blowout jets. By using the \sat{sdo} high resolution observations, a detailed analysis of a corona jet that underwent a standard-to-blowout transition has been reported by \cite{liu11}, of which an erupting blob preceded by a loop was considered to be the front-core structure of a bubble-like CME during its initiation stage on the disk observation. The authors thus speculated that the base arch of the jet underwent a miniature version of CME process. This study implied that bubble-like CME associated with the blowout eruption of the jet's base arch might be observed in the outer corona. However, it is a pity than the CME in \cite{liu11} did not be detected by any available coronagraph in the space or on the ground. Numerical simulations of the blowout of the twisted jet core field by the breakout reconnection with ambient fields were also conducted \citep{pari09,rach10}. Up to present, there is still no direct observational report on CMEs that are associated with the coronal blowout jets, even though \cite{moor10} have pointed out that the dichotomy of coronal jets might also be observed in the outer corona.

In this paper, for the first time, we present a coronal blowout jet that was associated a simultaneous bubble-like and a jet-like CMEs observed by the inner coronagraph (COR1, \cite{thom03}) on board {\sl Solar Terrestrial Relations Observatory} Ahead (\sat{stereo}, \cite{kais08}). The ejection occurred on 2011 July 22 close to the NOAA Active Region 11252 (N25W48). By using high temporal and spatial multi-wavelength and multi-angle observations taken by \sat{sdo}, \sat{stereo}, and Big Bear Solar Observatory (BBSO), we study the dynamics and evolutionary process in detail, and try to find a few new clues to understand the physics of the coronal blowout jets. In the rest of the paper, instruments and data sets are described in Section 2, results are presented in Section 3, interpretation, and discussions are given in Section 4, and conclusions are summarized in Section 5.

\section{INSTRUMENTS AND DATA SETS}
The full-disk H$\alpha$ center images are obtained from the BBSO, New Jersey Institute of Technology, USA. The H$\alpha$ observations have a cadence of 1 minute and a pixel width of $1\arcsec$. The {\sl SDO}/Atmospheric Imaging Assembly (AIA; \cite{leme11}) has high cadences up to 12 seconds and short exposures of 0.1--2 seconds. It captures images of the Sun's atmosphere out to \rsun{1.3}, with a pixel width of $0\arcsec.6$ in seven EUV and three UV-visible wavelength bands. {\sl SDO}/Helioseismic and Magnetic Imager (HMI; \cite{scho10}) provides full-disk line-of-sight magnetograms at a 45 seconds cadence with a precision of 10 G, and the pixel width is the same with AIA. While \sat{sdo} observes the Sun from the Earth viewpoint, the \sat{stereo} Ahead observes the Sun from the western side in another angle of view. On 2011 July 22, the separation angle of \sat{stereo} Ahead and \sat{sdo} was about $99^{\circ}$. The full-disk 195 \AA\ and 304 \AA\ images taken by the Extreme Ultraviolet Imager (EUVI; \cite{wuel04}) of the Sun Earth Connection Coronal and Heliospheric Investigation (SECCHI; \cite{howa08}) on board \sat{stereo} Ahead are used in this paper. The 195 (304) \AA\ images have a cadence of 2.5 (10) minutes and a pixel width of $1\arcsec.6$. The COR1 on board \sat{stereo} Ahead provides images at a 5 minutes cadence, and the field of view (FOV) is from 1.4 to \rsun{4}. The H$\alpha$ images are aligned with the \sat{sdo} images by using the sunspot of Active Region 11259 that was close to the disk center on 2011 July 22. In addition, all the images used in this paper are differentially rotated to a reference time (16:00:00 UT). Meanwhile, they are rotated such as north (east) is up (left).

\section{RESULTS}
\begin{figure}\epsscale{0.8}
\plotone{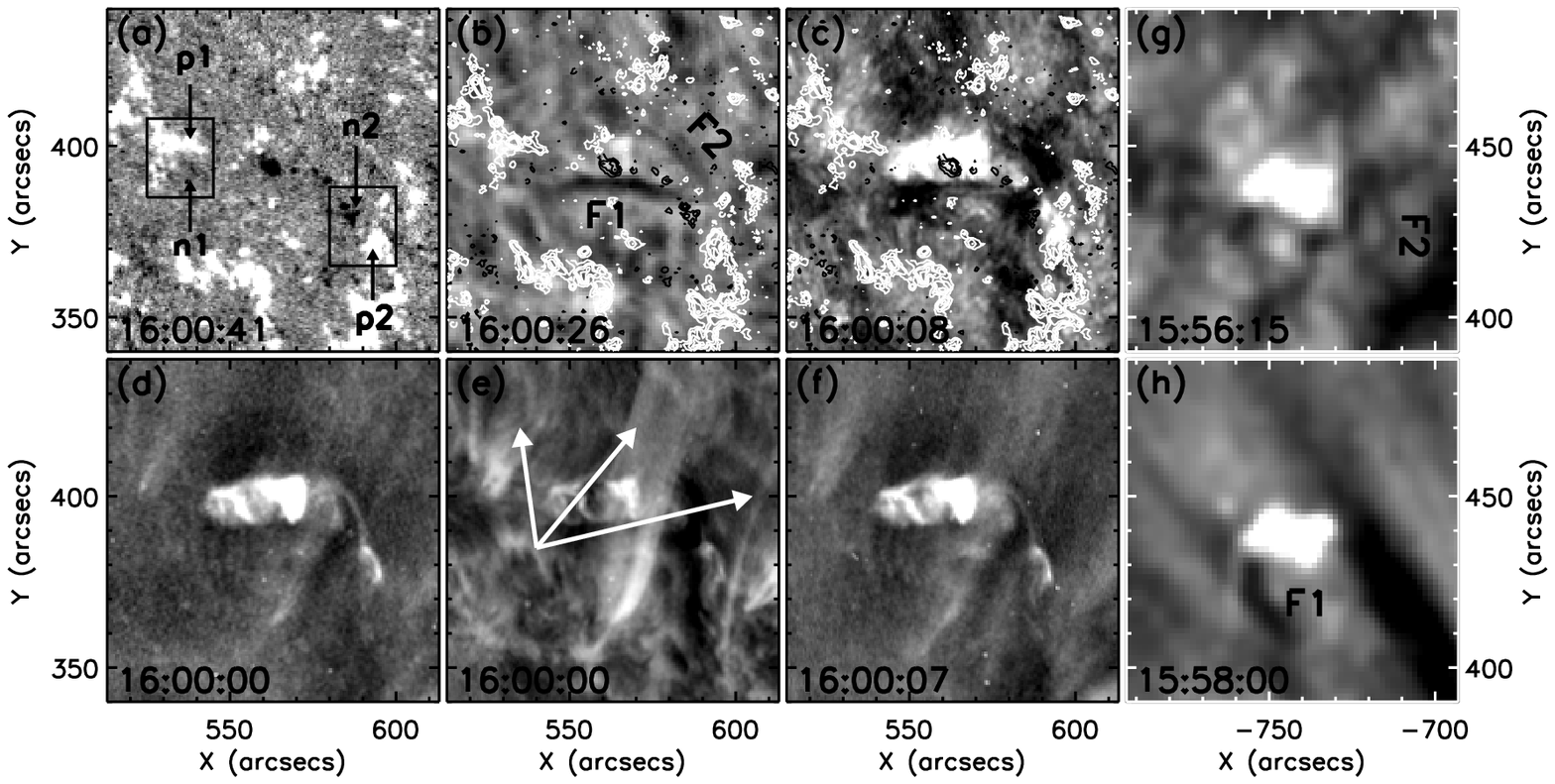} \caption{HMI line-of-sight magnetogram (a), BBSO H$\alpha$ center (b), AIA 304 \AA\ (c), AIA 211 \AA\ (d), AIA 171 \AA\ (e), AIA 193 \AA\ (f), {\sl STEREO} Ahead 304 \AA\ (g), and 195 \AA\ (h) images show the event before the eruptions. The contours of the HMI image at 16:00:41 UT are overlaid on panels (b) and (c). The contour levels are $\pm20$, $\pm50$, $\pm100$, and $\pm200$ G, with white (black) color for positive (negative) polarity. The jointed arrows in panel (e) point to some representative coronal loops. The two small filaments are labeled by F1 and F2 respectively. The FOV is $100\arcsec \times 100\arcsec$ for each panel. \label{overview}}
\end{figure}

The blowout jet occurred on 2011 July 22 close to the NOAA AR11252 (N25W48), and no significant soft X-ray flare was recorded by \sat{goes} during the ejection time interval. However, some small-scale energy releasing signatures could be detected from the observations. An overview of the event before the ejection is displayed in \fig{overview}. Panel (a) is an HMI magnetogram showing the magnetic configuration of the region where the jet would occur. The east black box indicates the bipolar magnetic region in which flux emergences, and cancellations were observed to be associated with the subsequent blowout jet. The polarities are labeled p1 and n1. Two filaments, labeled ``F1'' and ``F2'', could be seen clearly on the BBSO H$\alpha$ center and AIA 304 \AA\ images. On the \sat{stereo} Ahead images, the orientation of F1 was changed from the east-west to the southwest due to the different viewpoints of \sat{sdo} and \sat{stereo} Ahead (compare panels (b) and (h)). Meanwhile, F2 was looking like a thin curving thread on the western side of F1. The contours of the HMI light-of-sight magnetogram at 16:00:41 UT were overlaid on the BBSO H$\alpha$ and the AIA 304 \AA\ images, of which white (black) color represents the positive (negative) flux. It clearly shows that the two filaments were located on the two different neutral lines respectively. Obviously, the region of interest was dominated by the positive flux and mainly characterised by unipolar open fields (indicated by the jointed arrows in panel (e)). It is reasonable to imagine that there must have some closed loops connecting the opposite polarities and restricting the filaments to erupt, even thought they were ambiguous on the EUV images. In the following, we are going to analyze the blowout jet involving the eruption of F1 in detail.

\fig{eruption1} shows the evolutionary process of the blowout jet on the H$\alpha$ center, AIA 304 \AA\ and 193 \AA\ time sequence images. The outer profile of F1 detected from the H$\alpha$ image at 16:13:58 UT is overlaid on the HMI line-of-sight magnetogram, and the relevant small bipolar magnetic region was magnified and inserted in the same panel (panel (a)). At around 16:13 UT, a conspicuous bright patch appeared around the interface of p1 and n1, and it could be identified at the chromospheric and EUV lines (indicated by the short vertical white arrows in \fig{eruption1}). It is interested that the appearance of the bright patch on the AIA 304 \AA\ (or H$\alpha$) images was earlier than that on the AIA 193 \AA\ images. This suggests that the bright patch may result from a small-scale reconnection event in the lower atmosphere. About four minutes later, F1 began to rise slowly and the region surrounding F1 was ignited entirely around 16:24 UT except the southwest side, which suggested that the field lines striding F1 were pushing up and thereby reconnecting with the surrounding open fields. In addition, the rising F1 showed obvious untwisting motions during the rising period. At around 16:27 UT, the jet could be distinguished, and the two flare ribbons were also obvious on either side of the original site of F1. It is interested that the jet showed simultaneous hot and cool components next to each other during the developing stage. The two components of the blowout jet could be observed not only at EUV wavelength bands but also at H$\alpha$ center. By comparing the jet at different wavelength bands carefully, we find that the two components were cospatial and held a similar evolutionary pace (see the fifths column of \fig{eruption1}). After 16:38 UT, the hot component disappeared and the jet showed as a dark and collimating feature that was composed of many thin dark threads, and the post-eruptive loops were conspicuous next to the footpoint of the jet body. The disappearance of the hot component was possibly because of the cooling of this component at this time. In addition, by seeing the time-elapse movie of various wavelength bands (video1.mpg; available in the online journal), we find that the cool component was directly evolved from the erupting F1, while the hot component was resulted from the outward moving heated plasma, which was generated by the reconnection between the loops striding F1 and its ambient opposite-polarity open field lines. We note that the eruption of F1 just resembled the evolutionary process of a large-scale filament eruption that is often tightly associated with large-scale CME. Furthermore, lateral motions of the jet body and unwinding motions around the main axis of the jet was also observed, which manifested the releasing of the magnetic helicity stored in the sheared core field, as in the unwinding polar jet presented by \cite{shen11b}. The unwinding direction was counterclockwise when looking from above along the jet axis. The apparent speed of the jet top in the plane of the sky was about \speed{235}, which is obtained by tracing the front of the jet on the successive AIA 304 \AA\ images.

The positive and negative magnetic fluxes of the box region shown in the \fig{eruption1} (a), and light curves of the same region (also the region of the bright patch) at various wavelength bands are measured from the successive observations. The results are shown in \fig{flux1}. We divide the changing of the fluxes into two stages, viz 16:06 UT -- 16:16 UT (before the ejection), and 16:16:UT -- 16:30 UT (during the ejection). For the first stage, the positive flux was decreasing at the beginning (before 16:09 UT), but then it was increasing quickly to a higher value. On the other hand, the negative flux first showed a rapid increasing phase during the decreasing period of the positive flux, but then the increased flux was reduced to the initial level (16:09 UT -- 16:12 UT). Finally, the negative flux also increased to a higher value. Such changing pattern of the fluxes suggests that the emerging of the negative flux and the cancellation between the opposite polarities had occurred alternately. The cancellation can also be observed on the HMI light-of-sight successive images. In the meantime, all the light curves of the same region at various wavelength bands showed a moderate increase during the first stage, which indicated the appearance of the bright patch observed on the chromospheric and EUV images. It is important to note that the start time of this increase is the same with the beginning time of the negative flux's increase, which indicated that the bright patch was possibly resulted from the cancellation of the positive flux and emerging negative flux. In addition, the increase of the chromospheric 304 \AA\ line (\ion{He}{2}; $\log T=4.7$), which images the upper chromosphere of the Sun, was slightly earlier than that of the other coronal EUV lines. Since the magnetic cancellations on the photosphere is often considered as the manifestation of the magnetic reconnection in the lower atmosphere \citep{chae98,chae99}, we suggest that the altitude of site of the reconnection that was related to the cancellation discussed above was under the transition region, or the upper chromosphere. For the second stage, the positive flux showed a roughly linear decrease, while the negative flux first decreased but then increased to a higher level. This flux changing pattern indicated that the impulsive cancellation between the opposite polarities was occurred during the ejection of the blowout jet. In addition, the light curves also showed a violent increase during this period.

\begin{figure}\epsscale{0.8}
\plotone{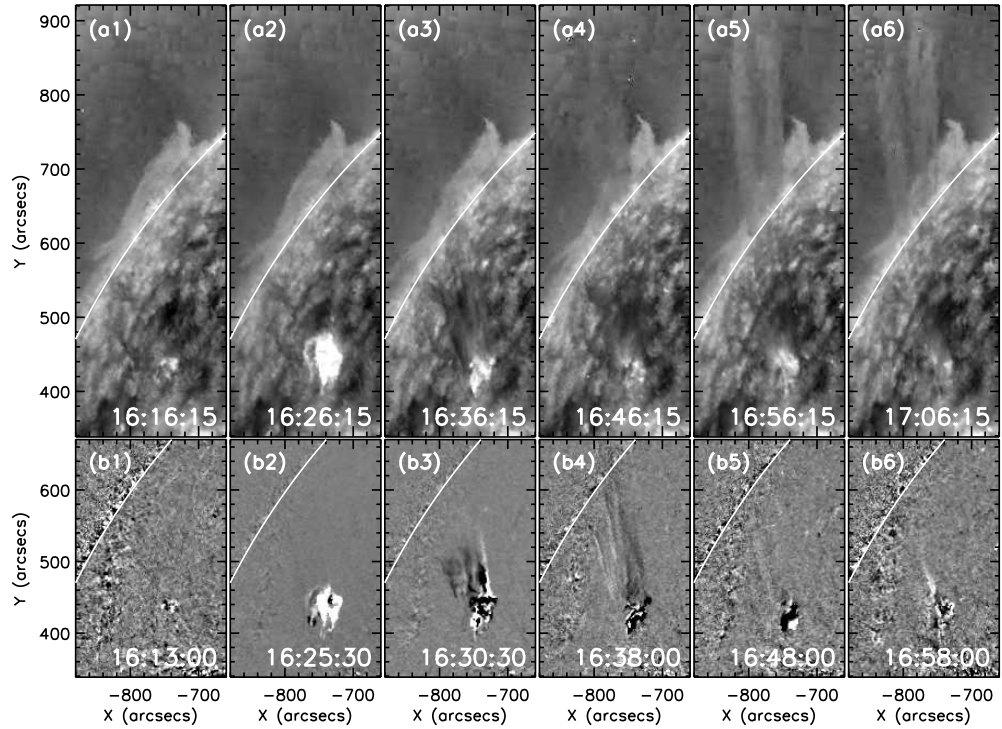} \caption{{\sl STEREO} Ahead 304 \AA\ direct (a1)--(a6) and 195 \AA\ difference (b1)--(b6) images show the coronal blowout jet. The white curve in each panel marks the disk limb. The FOV is $220\arcsec \times 680\arcsec$ for panels (a1)--(a6), and $220\arcsec \times 330\arcsec$ for panels (b1)--(b6). \label{stereo}}
\end{figure}

From the \sat{stereo} Ahead viewpoint, the blowout jet showed different intriguing characteristics. The results are shown on the 304 \AA\ direct and 195 \AA\ difference images \fig{stereo}. The characteristics such as the bright patch close to the jet base before the jet's initiation, simultaneous cool and hot components next to each other, and the two flare ribbons could also be identified. Due to the projection effect from different viewpoints, the hot component of the jet on the \sat{stereo} Ahead images was located on the western of the cool component, which was opposite to the location that was observed on the \sat{sdo} (see panel (b3)). An intriguing characteristic of the blowout jet on the \sat{stereo} Ahead images was the bifurcation of the jet body when it was propagating out. This phenomenon made the jet appeared to be composed of two separate branches of plasma flows (see panels (a5) and (a6)). On the basis of the spatial relationship between the cool and the hot components of the jet, we conjecture that the jet on the \sat{stereo} Ahead 304 \AA\ images was possibly just the cool component, which was resulted from the erupting F1. In addition, by checking the evolutionary process of the jet on the \sat{stereo} and \sat{sdo} successive observations, we identified that the bifurcation of the jet body was resulted from the uncoupling motions of the F1's two legs that were highly twisted at the very beginning. The bifurcation of the jet body did not be observed on the \sat{sdo} images, which was possibly due to the projection effect from different viewpoints, since the two legs of F1 were likely overlapping together when looking from the \sat{sdo} viewpoint.

\begin{figure}\epsscale{0.8}
\plotone{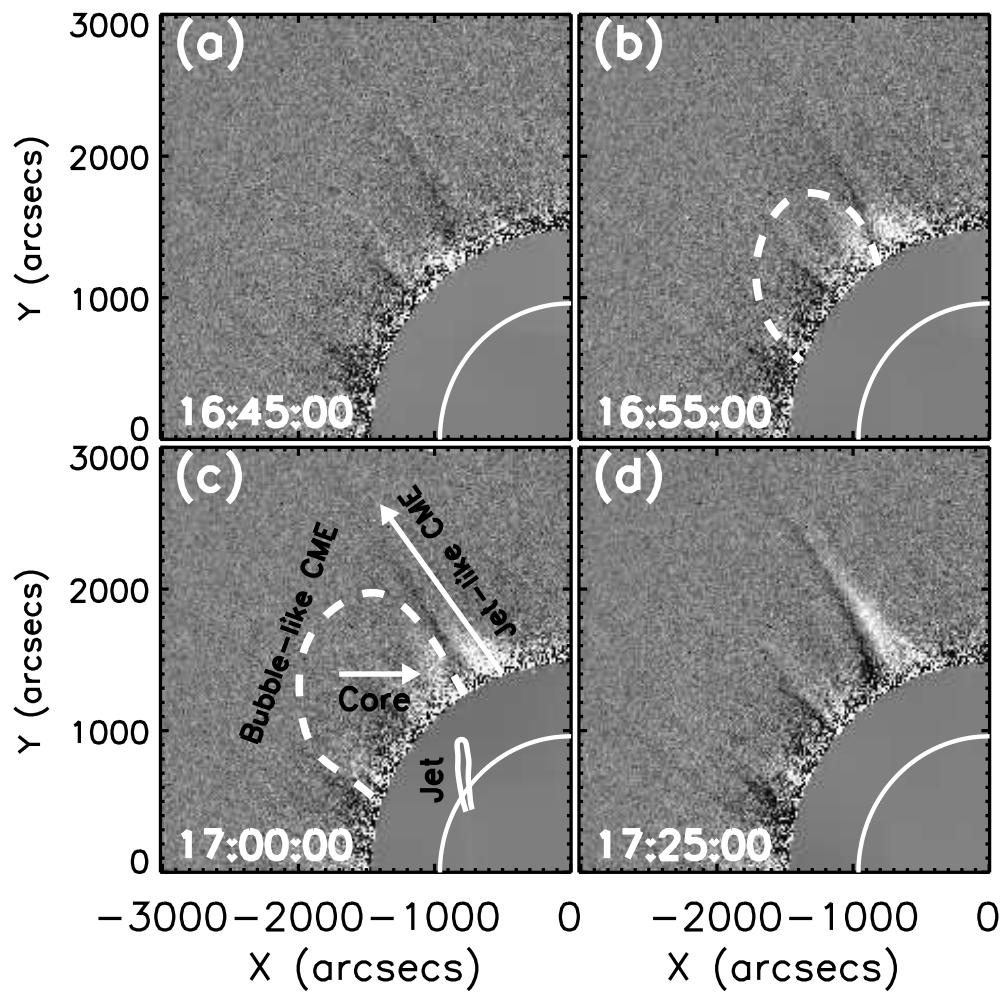} \caption{{\sl STEREO} Ahead COR1 fixed-base difference images show the simultaneous bubble-like and the jet-like CMEs. The dashed curves mark the bright front of the bubble-like CME, while the long white arrow indicates the jet-like CME. The short horizontal arrow points to the bright core of the bubble-like CME. The profile of the blowout jet measured from the \sat{stereo} Ahead 304 \AA\ image at 16:56:15 UT is overlaid in panel (c). The FOV for each panel is $3000\arcsec \times 3000\arcsec$. \label{cme}}
\end{figure}

It is striking that a simultaneous bubble-like and a jet-like CMEs were observed to associate with the blowout jet in the FOV of the COR1 Ahead. The CMEs are displayed in \fig{cme}. The jet-like CME first reached the FOV of the COR1 at around 16:45 UT, and then followed by the bubble-like CME on the eastern side of the jet-like CME. Since the bubble-like CME was very faint on the COR1 images, we use a dashed white curve to mark the bright front of the bubble-like CME, while the jet-like CME was also indicated by the long white arrow. In addition, a time-lapse move (video2.mpg) could be found in the online version of the journal, from which one can resolve the two simultaneous CMEs easily. For the bubble-like CME, it showed a typical tree-part structure, with a bright core preceded by a dark cavity and a bright front. The bright core, which often represents the erupting filament in large-scale CME eruptions, was located near the western edge of the bright CME front rather than at the center as usual. However, it was consistent with the direction of the blowout jet's cool component that has been identified as the erupting F1 (see panel (c) of \fig{cme}). The close temporal and spatial relationship between the bubble-like CME and the erupting F1 suggested that this bubble-like CME was directly caused by the eruption of F1 via a similar mechanism in larger filament-CME eruptions, even though the eruption of F1 was involved in the coronal blowout jet. For the jet-like CME, its direction was in agreement with the hot component of the blowout jet. So we propose that this jet-like CME was just the outward extension of the blowout jet's hot component, as the jet-like CMEs (or white-light jets) documented in the previous statistical studies \citep{wang00,wang02}. It seems that the two simultaneous CMEs were independent events but dynamically correlated to each other.

\section{INTERPRETATION AND DISCUSSION}
\begin{figure}\epsscale{0.8}
\plotone{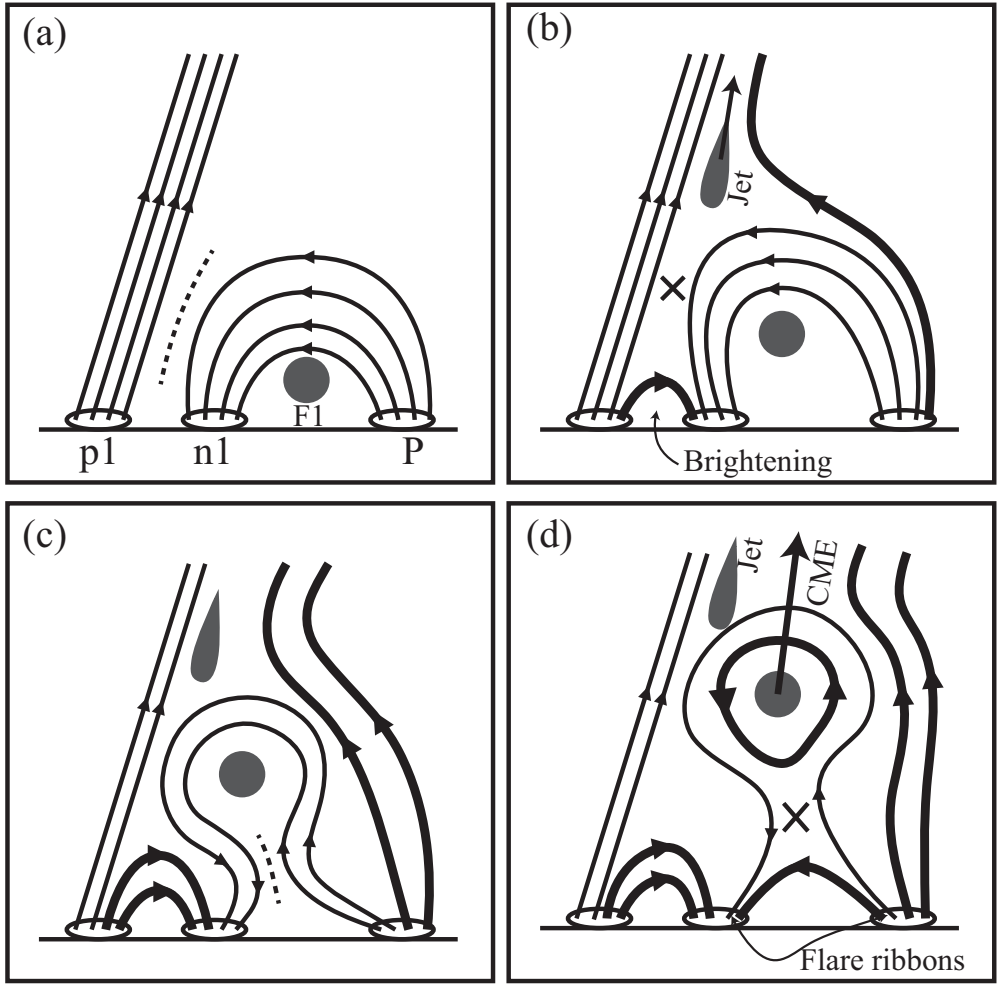} \caption{Cartoon demonstrating the coronal blowout jet and the associated simultaneous CMEs. (a) the initial magnetic configuration. (b) the external reconnection and the generation of the hot component of the jet that causes the jet-like CME observed. (c) the new current sheet beneath F1. (d) the internal reconnection and the production of the bubble-like CME. The dashed lines indicate the current sheets, and the thick lines represent the reconnected field lines. The reconnection sites are labeled by the ``X'' symbols. \label{cart}}
\end{figure}

Based on our analysis results, we interpret the physical mechanism of the coronal blowout jet and the associated bubble-like and the jet-like CMEs recorded by the COR1 on board \sat{stereo} Ahead, with a two-dimensional cartoon at four critical moments (\fig{cart}). In the cartoon, only a few representative field lines are drawn. The primary magnetic configuration is shown in panel (a), of which a closed loop system connecting n1 and P (the positive polarity on the northern of n1) is close to a bundle of unipolar open field lines, and a current sheet between them can be expected (the dashed line). It should be noted that the closed loop system is assumed to be sheared, within which is the small filament (F1). In our observation, the bright patch before the initiation of the jet was close to the jet base, and it could be observed simultaneously at chromospheric and EUV lines. It was closely related with the flux cancellation observed on the HMI line-of-sight magnetograms, and we explain these phenomena as the manifestation of the reconnection in the lower atmosphere on the photosphere. Panel (b) shows the external reconnection between the closed field and the open field lines. The reconnection will yield three main observable effects: the brightening below the reconnection site, the upward hot component of the jet, and the rising of F1 and its overlying closed fields. The first observable effect is the manifestation of the downward closed, heated reconnected loops, which is expected to be observed as a microflare at chromospheric and coronal EUV lines, and flux cancellation on magnetograms. The outward moving heated plasma is expected to be observed as a bright jet on the disk observations or a jet-like CME in the outer corona. The external reconnection continuously removes the field lines that are restricting F1 to erupt and thereby reduces the confining capacity of these fields. Thus F1 is going to rise shortly after the start of the reconnection. On the other hand, a new current sheet will form between the two legs of the confining field lines underneath F1 during this rising period (dashed line in panel (c)). When the new formed current sheet gets thin enough, the internal reconnection within it will start and thus lead to the eruption of F1, two flare ribbons on either side of the original site of F1, and a bobble-like CME observed in the outer corona (panel (d)).

It should be noted that the cartoon model presented here is just a combination of the standard jet model \citep[e.g.,][]{shib92} and the classical flux rope eruption model \citep[e.g.,][]{lin00}. The primary mechanism is similar to the so-called magnetic breakout model \citep{anti99,shen11c}, where the external reconnection transfers flux that tends to hold down the sheared low-lying field to the lateral flux systems and thereby allows the sheared core field to erupt explosively outwards. The model can also be viewed as an extension of the model of the first type blowout jet proposed by \cite{moor10}. In our blowout jet model, many observable characteristics of the blowout coronal jets, which involves a filament eruption within base arch, can be explained naturally, especially the simultaneous bubble-like and the jet-like CMEs. The blowout jet presented in this paper and the associated model suggests that some different kinds of large-scale solar activities can occur simultaneously and integrate into one solar eruption event, even though they abide by different physical mechanisms.

The bifurcation of the jet body on the \sat{stereo} 304 \AA\ images indicated that it was resulted from the untwisting motions of the erupting filament's two legs that were twisted at the very beginning. Such bifurcation phenomenon of coronal jet has also been reported previously by \cite{alex99}, they interpreted it as a dynamic response to the rapid transport of twist along the field, or the development of an instability as the jet propagates out. Our observational results are basically in agreement their first explanation. On the other hand, the traditional concept of the corona jets is possibly just the hot component of the jet observed in the present case \citep{shib92}. The analysis results indicate that the cool component was, in fact, the erupting filament, which is a new concept about the jet phenomena.

In the previous studies, \cite{chae99} found that hot and cool ejections on H$\alpha$ images, the cool jet-like feature on H$\alpha$ images was cospatial and simultaneous with the EUV coronal jets, but the hot H$\alpha$ ejection was not. We conjecture that the hot ejection did not be observed at EUV wavelength bands was probably due to the lower cadence of the observations that they have used, since the hot ejecta might have been cooled down when the imager captures the next image. In other studies \citep[e.g.,][]{alex99,jian07}, the cool and hot components were also observed, and the appearances of the cool components were found to delay the hot components several minutes. The authors explained that such a delay might be caused by the cooling of the earlier, hotter jet material. Thanks to the excellent observations taken by \sat{sdo}, \sat{stereo}, and BBSO, we can successfully distinguish the progenitors of the simultaneous cool and the hot components of the coronal blowout jet in the present event, of which the cool component was evolved from the filament contained within the jet's base arch, while the hot component was the outward moving heated plasma generated by the reconnection between the closed fields that confined the filament and their ambient open field lines. In our model (\fig{cart}), there will be a simultaneous bubble-like and a jet-like CMEs associated with the coronal blowout jets. Observationally, such bubble-like CMEs are usually so weak that some authors could not find them \citep[e.g.,][]{moor10,liu11}. Anyway, more similar observational studies rely on high-resolution multi-wavelength and multi-angle observation in a wide range of temperature are desired in the future.

\section{CONCLUSION}
Combining the analysis of high temporal and high spatial resolution multi-wavelength observations from two different viewpoints, we present a coronal blowout jet and a mini-filament eruption to learn the trigger mechanism of them. To our knowledge, this is the first report that a simultaneous bubble-like and a jet-like CMEs were dynamically related to one coronal blowout jet. Our main analysis results are summarized as follows. (1) A bright patch at various lines slightly preceding the corresponding flux cancellation on the photosphere were observed at the same location that was close to the footpoint of the blowout jet. The start of the brightening was occurred a few minutes before the initiation of the blowout jet, and the light curves further showed that the brightening at the chromospheric line was slightly earlier than that at the coronal EUV lines. We interpret these phenomena as the manifestations of the external reconnection of the blowout jet in the lower atmosphere, and the reconnection site was possibly under the transition region, or the upper chromosphere. (2) The blowout jet showed simultaneous hot and cool components next to each other, and they could be observed in a wide temperature range. The hot component was identified as the outward moving heated plasma generated by the external reconnection, while the cool component was the erupting filament that was contained within the sheared base arch. (3) From the \sat{stereo} Ahead viewpoint, the jet experienced a bifurcation process as it was propagating out. This phenomenon was resulted from the untwisting motions of the erupting filament. The two bifurcated branches were identified as the erupting filament's two legs, which was tightly twisted at the very beginning. (4) A simultaneous bubble-like CME involving a three-part structure and a jet-like CME were evidenced in the FOV of the COR1 Ahead. We find that the bubble-like CME was associated with the eruption of filament eruption contained within the base arch (also the cool component of the jet), while the jet-like CME was the extension of the hot component of the coronal blowout jet in the outer corona. For figuring out the question whether such a simultaneous bubble-like and a jet-like CMEs associating with one coronal blowout jet is a common phenomenon or not, more similar observational studies of coronal blowout jets are desired in the future.

\begin{figure}\epsscale{0.8}
\plotone{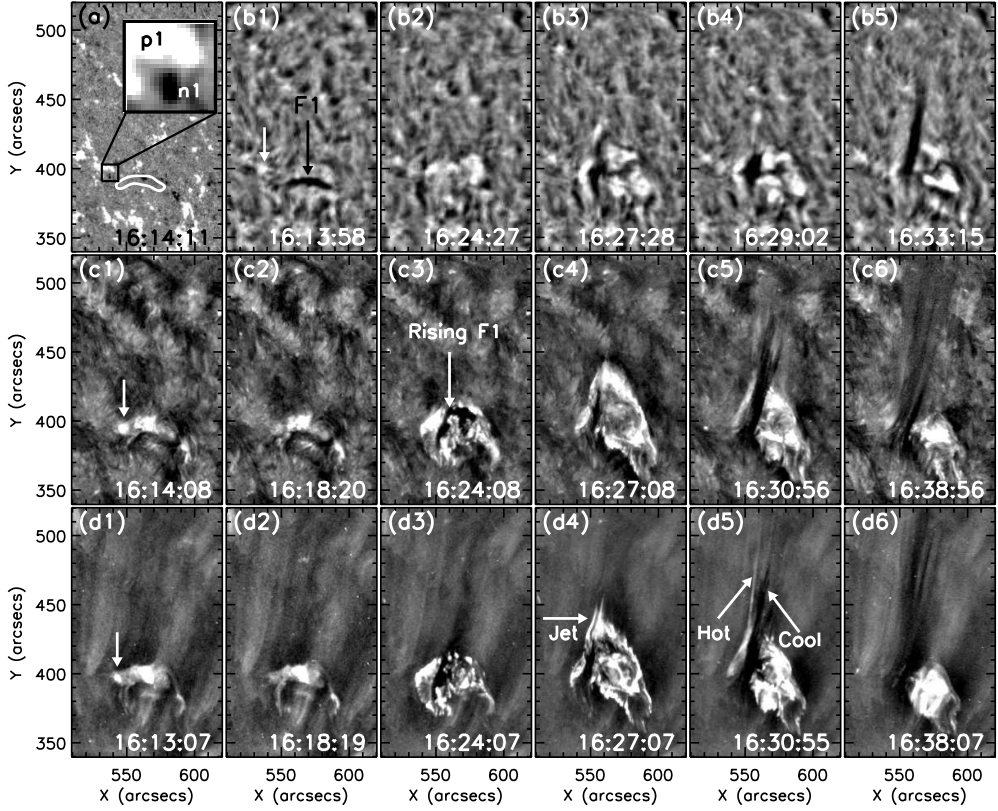} \caption{Time sequences of BBSO H$\alpha$ center (b1)--(b5), AIA 304 \AA\ (c1)--(c6), and AIA 193 \AA\ (d1)--(d6) images show the blowout jet and the associated filament at the base. Panel (a) is an HMI line-of-sight magnetogram just prior to the initiation of the jet, and the white contour is the profile of F1. The inserted image in panel (a) is the magnified magnetic region of the small rectangle region. The vertical short arrows indicate the bright patch before the ejection, while the two white arrows in panel (d5) point to the hot and cool components of the ejecting jet. The FOV for each frame is $110\arcsec \times 180\arcsec$. \label{eruption1}}
\end{figure}

\begin{figure}\epsscale{0.5}
\plotone{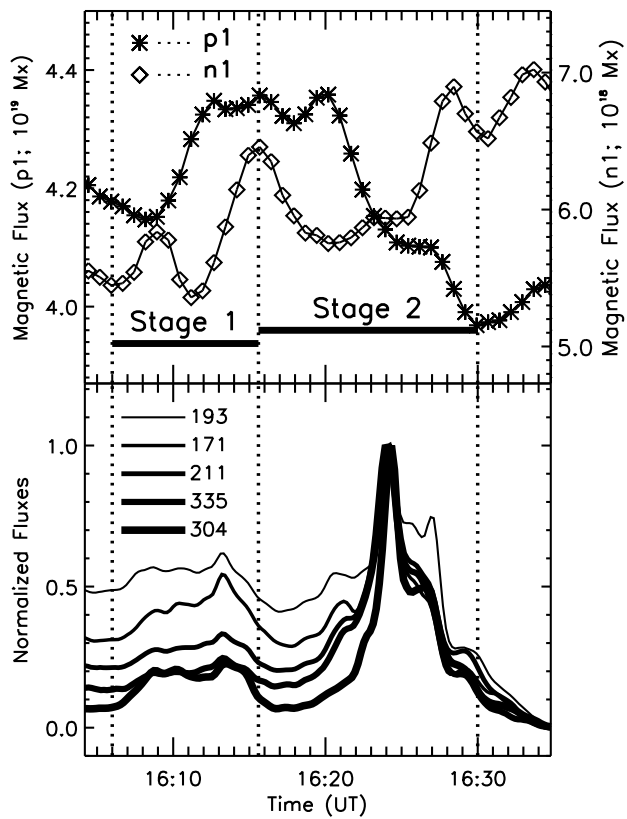} \caption{Top panel shows the positive (asterisk) and negative (diamond) magnetic fluxes within the black box shown in panel (a) of \fig{eruption1} during the blowout jet period, of which the absolute value of the negative flux is scaled by the y-axis on the right. Bottom panel shows the light curves of various wavelength bands. They are measured from the same region (also the region of the bright patch), and displayed in an arbitrary unit to fit in the panel. The three vertical dashed lines indicate changing stages of the magnetic fluxes. \label{flux1}}
\end{figure}

\begin{figure}\epsscale{0.8}
\plotone{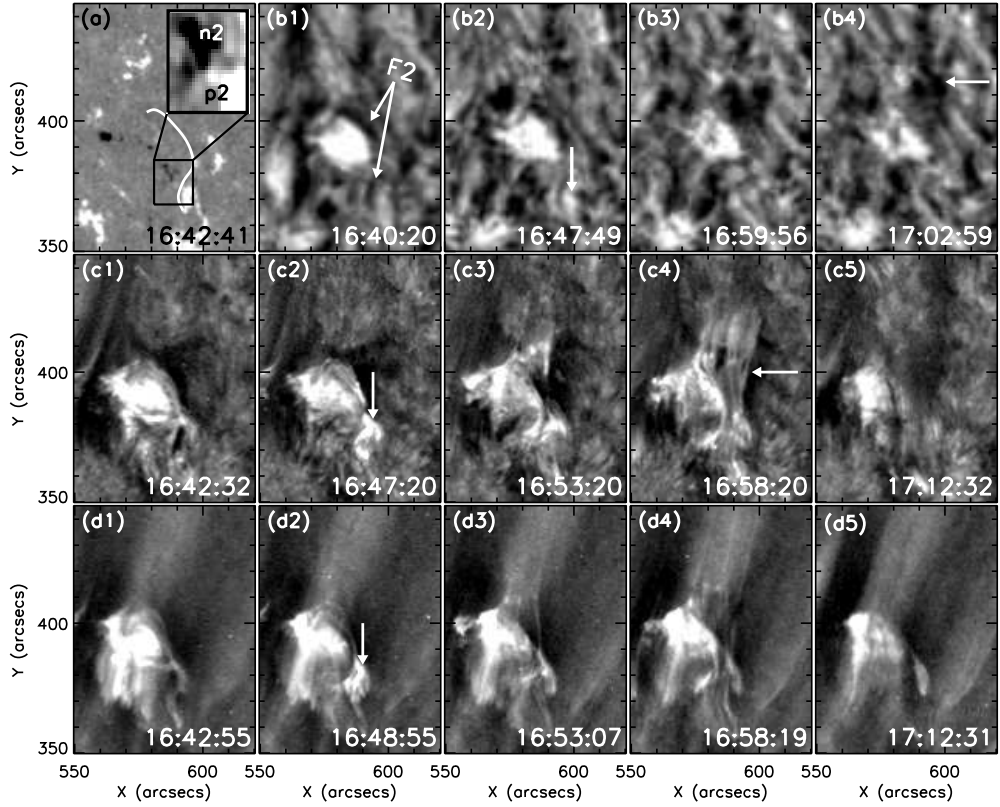} \caption{Time sequences of BBSO H$\alpha$ center (b1)--(b5), AIA 304 \AA\ (c1)--(c6), and AIA 193 \AA\ (d1)--(d6) images show the eruption of F2. Panel (a) is an line-of-sight magnetogram overlaid by the spine of F2 (white curve), which is detected form the H$\alpha$ image at 16:40:20 UT. The magnetic region of p2 and n2 (shown in \fig{overview} (a)) is magnified and inserted in the same image. The vertical short arrows point to the bright patch, while the horizontal arrows indicate the erupting F2. The FOV is $70\arcsec \times 95\arcsec$ for all panels. An animation of this figure is available in the online journal (video3.mpg). \label{eruption2}.}
\end{figure}

Following the blowout jet presented thereinbefore, an inverse S-shaped mini-filament (F2) erupted on the jet's western side at about 16:50 UT. No obvious observable stamp that could relate the two successive eruptions. The detailed eruption process is shown in \fig{eruption2}. Similar to \fig{eruption1}, the magnetic region of interest just before the activation of F2 was magnified and inserted in the same panel (panel (a)). It is clear that the positive (p2) and the negative (n2) polarities were close to each other. Flux convergence and cancellation between p2 and n2 were observed on the HMI magnetograms. At about 16:40 UT, F2 had developed into an inverse S-shape dark filament, it was conspicuous on the H$\alpha$ image but ambiguous at EUV lines. At about 16:47 UT, an obvious bright patch around the interface of p2 and n2 was observed on the H$\alpha$ center and the EUV images (see the vertical white arrows in panels (b2), (c2) and (d2) in \fig{eruption2}). It was overlapped with the southern segment of F2 and shown as the two flare ribbons of a microflare. This result might imply that the bright patch was possibly resulted from the tether-cutting reconnection of the legs of the field lines that were confining F2 \citep{moor01}. Subsequently, F2 started to erupt violently as a corona jet (see the horizontal white arrows in \fig{eruption2}), and unwinding motions was also observed. The unwinding direction was the same with the blowout jet, which was in agreement with the so-called hemispheric helicity rule that patterns of negative helicity occur predominantly in the northern solar hemisphere, and those of positive helicity in the south \citep{seeh90}. The apparent speed in the plane of the sky of the eruption of F2 was about \speed{48}.

It should be noted that this mini-filament eruption was analogous to the initiation of the other type of coronal blowout jets proposed by \cite{moor10}, of which the sheared core field starts erupting on its own before any external breakout reconnection. Here F2 did erupt first on its own because of the possibly reconnection underneath it, but no specific evidence for the external reconnection could be identified from the multi-wavelength observations. In addition, the magnetic environments surrounding F2 were also not favorable for the external reconnection to occur. This event suggests that a lots of jet-like solar explosive events are possibly, in fact, mini-filament eruptions.

Previous studies have shown that a lots of mini-filament eruptions are usually taken the shape of coronal jets. They were found to associate with cancellations and convergences of opposite fluxes, and accompanied by flare-like brightenings, post-eruptive flare loops, coronal dimmings, and mini-CMEs \citep[e.g.,][]{wang00,ren08,hong11}. Occasionally, they were observed to be failed from escaping the Sun \citep[e.g.,][]{liuy09,shen11a}. In our case, no CME could be found to be associated with the mini-filament eruption, which was possibly due to the low erupting speed of the filament. Here, we appeal for more attentions to be paid to distinguishing whether a large number of jet-like eruptive solar events are coronal jets or filament eruptions, which rests upon high temporal and hight spatial resolution multi-wavelength observations.

\acknowledgments \sat{sdo} is a mission for NASA's Living With a Star (LWS) Program. The \sat{stereo}/SECCHI data used here were produced by an international consortium of the Naval Research Laboratory (USA), Lockheed Martin Solar and Astrophysics Lab (USA), NASA Goddard Space Flight Center (USA), Rutherford Appleton Laboratory (UK), University of Birmingham (UK), Max-Planck-Institut for Solar System Research (Germany), Centre Spatiale de Li$\rm \grave{e}$ge (Belgium), Institut d'Optique Th$\rm \acute{e}$orique et Appliqu$\rm \acute{e}$e (France), and Institut d'Astrophysique Spatiale (France). The H$\alpha$ observations are obtained from the Big Bear Solar Observatory (BBSO), New Jersey Institute of Technology, USA. We thank an anonymous referee for his/here helpful comments and valuable suggestions. This work is supported by the Natural Science Foundation of China (Grant Nos. 10933003, 11078004, and 11073050), MOST (2011CB811400), and Open Research Program of Key Laboratory of Solar Activity of National Astronomical Observatories of Chinese Academy of Sciences (KLSA2011\_14).

\end{document}